\documentclass[
 reprint,
superscriptaddress,
nofootinbib,
nobibnotes,
 amsmath,amssymb,
prl,
]{revtex4-1}
\usepackage{graphicx}
\usepackage{dcolumn}
\usepackage{bm}
\usepackage{notoccite}
\usepackage{physics}
\usepackage{xcolor}
\usepackage{mathtools}
\begin{document}

\title{Tuning topology in thin films of topological insulators by strain gradients}

\author{Raffaele Battilomo}
\affiliation{Institute for Theoretical Physics, Center for Extreme Matter and Emergent Phenomena, Utrecht University, Princetonplein 5, 3584 CC Utrecht,  Netherlands}
\author{Niccol\'o Scopigno}
\affiliation{Institute for Theoretical Physics, Center for Extreme Matter and Emergent Phenomena, Utrecht University, Princetonplein 5, 3584 CC Utrecht,  Netherlands}
\author{Carmine Ortix}
\affiliation{Institute for Theoretical Physics, Center for Extreme Matter and Emergent Phenomena, Utrecht University, Princetonplein 5, 3584 CC Utrecht,  Netherlands}
\affiliation{Dipartimento di Fisica ``E. R. Caianiello", Universit\'a di Salerno, IT-84084 Fisciano, Italy}

\begin{abstract}
We theoretically show that the coupling of inhomogeneous strains to the Dirac fermions of three-dimensional topological insulators (3DTI)  in thin film geometries results in the occurrence of phase transitions between topologically distinct insulating phases. By means of minimal ${\bf k \cdot p}$ models for strong 3DTI in the Bi$_2$Se$_3$ materials class, we find that in thin films of stoichiometric materials a strain-gradient induced structure inversion asymmetry drives a phase transition from a quantum spin-Hall phase to a topologically trivial insulating phase. Interestingly, in alloys with strongly reduced bulk band gaps strain gradients have an opposite effect and promote a topologically non-trivial phase from a parent normal band insulator. 
These strain-gradient assisted switchings between topologically distinct phases are expected to yield a flexomagnetic coupling in magnetic topological insulator thin films. 
\end{abstract}

\maketitle

\paragraph{Introduction --}
In condensed matter systems characterized by low-energy Dirac or Weyl quasiparticles, the coupling between electronic and lattice degrees of freedom leads to remarkable effects. In graphene~\cite{cas09}, for instance, elastic deformations of the lattice due to strains couple to the Dirac fermions as a pseudomagnetic vector potential~\cite{guinea2009,vozmediano2010,amorim2016}. Spatially inhomogeneous strains then typically result in a pseudomagnetic field, which, in its simplest form, gives rise to a pseudo-Landau level spectrum at low energies~\cite{guinea2009}. Such strain-induced pseudo-Landau levels have been observed using scanning tunneling microscopy techniques both in graphene~\cite{Levy2010} and in artificial graphene~\cite{gomes2012}. When the pseudomagnetic fields are spatially nonuniform and periodically modulated~\cite{venderbos2016}, electron-electron interactions can trigger an integer quantum Hall effect without any external time-reversal-breaking field and thus similar in nature to the Haldane phase~\cite{Haldane1988}.  

In three-dimensional topological semimetals with Weyl quasiparticles~\cite{Armitage2018} instead, pseudomagnetic fields due to inhomogeneous strains~\cite{cortijo2015} have been predicted to yield an enhancement of the bulk conductivity due to an underlying chiral pseudomagnetic effect~\cite{Grushin2016}. Moreover, time-dependent mechanical 
deformations due to, {\it e.g.}, acoustic waves create a related pseudoelectric field~\cite{Pikulin2016}. The concomitant presence of pseudoelectric and pseudomagnetic fields can also lead to a pseudo-Landau level collapse~\cite{Arjona2017}, which generalizes the Landau level collapse that occurs when perpendicular ordinary electric and magnetic fields are present.  

Low-energy Dirac quasiparticles are also naturally realized at the boundaries of time-reversal invariant strong three-dimensional topological insulators (3DTIs)~\cite{hasan2010,qi2011}. Contrary to the Dirac states of graphene, however, these surface states are anomalous: on each separate surface they violate the fermion doubling theorem~\cite{NIELSEN1983} and thus form half an ordinary  two-dimensional (semi)metal. This apparent paradox is resolved by considering that the Dirac states appearing at the opposite surface cancel the anomaly and hence regularize the system.

A 3DTI thin film then can be thought of as forming a graphene analog but where the Dirac quasiparticles are separated in real space rather than momentum space, and acquire a thickness-dependent mass due to the symmetry-allowed tunneling effect between opposite surfaces. Moreover, these geometries represent the ideal playground where inhomogeneous strain effects become pervasive. Thin films are much more flexible than bulk crystals and can bend easily~\cite{Schmidt2001}. In conventional semiconducting systems, 
substantial bending strain effects have been both theoretically predicted~\cite{ortix2011} and experimentally observed~\cite{mei2007}. Additionally, in thin films grown on a substrate a lattice mismatch at the interface yields a strain that gradually relaxes away from the interface thus providing yet another source of non uniform strains. 

Starting out from these observations, in this Letter we show that already in their simplest form inhomogeneous strains couple to the Dirac fermions of 3DTI thin films 
in an entirely different manner as compared to graphene and topological semimetals. Bending 
strains
trigger indeed a phase transition between two topologically distinct insulating phases. 
In materials with large bulk band gaps, strain gradients result in a structure inversion asymmetry term driving a topological phase transition from a quantum spin-Hall (QSH) phase to a conventional band insulator~\cite{Shan2010}. Remarkably, we find that in materials with strongly reduced bulk band gaps 
inhomogeneous strains 
have an opposite effect, and promote a topologically non-trivial QSH phase thanks to a strain-gradient analog of the quantum-confined Stark effect~\cite{cendula2012}.

\paragraph{Strained TI thin films --}
In order to analyze the influence of inhomogeneous strains in thin films of strong 3DTIs, we start out by introducing the effective bulk ${\bf k} \cdot {\bf p}$ Hamiltonian close to the $\Gamma$ point of the Brillouin zone (BZ) for the Bi$_2$Se$_3$ family of materials. It can be derived using the theory of invariants which accounts for the essential point group symmetries of the $R{\bar 3}m$ (No. 166) space group, namely the inversion symmetry ${\mathcal I}$ and the three-fold rotation symmetry along the $z$-axis ${\mathcal C}_{3 z}$, with the addition of time-reversal symmetry ${\mathcal T}$. 
Using as basis states the $\Gamma$-point spin-orbit coupled parity eigenstates $\ket{P1^+_z,\frac{1}{2}},\ket{P2^-_z,\frac{1}{2}},\ket{P1^+_z,-\frac{1}{2}},\ket{P2^-_z,-\frac{1}{2}}$, which are derived from the hybridized $p_z$ orbitals of Bi and Se, the three  symmetries can be represented as  $\mathcal{T}=i(\sigma_2\otimes\tau_0)\mathcal{K}$, $\mathcal{I}=\sigma_0\otimes\tau_3$ and ${\mathcal C}_{3z}=\exp(i(\pi/3)\sigma_3\otimes\tau_0)$, with $\mathcal{K}$ being the complex conjugation, and $\sigma$ and $\tau$  the Pauli matrices acting in the spin and orbital space, respectively. With this, the low-energy continuum bulk Hamiltonian~\cite{Zhang2009} can be cast in the following form
\begin{multline}
{\mathcal H}_{\mathrm{3DTI}}(\mathbf{k}) = \epsilon_0 ({\bf k}) \, \sigma_0 \otimes \tau_0 + (-M+B_1 k^2_z+B_2 k^2_\parallel) \sigma_0\otimes \tau_3 \\+A_1 k_z \sigma_3\otimes\tau_1+A_2 k_x \sigma_1\otimes\tau_1 +A_2 k_y \sigma_2\otimes\tau_1,
\label{H0}
\end{multline}
where $\epsilon_0({\bf k})=D_1 k_z^2 + D_2 k_{\parallel}^2$ and $k_\parallel^2=k_x^2+k_y^2$. In addition, $A_{1,2}$, $B_{1,2}$, $D_{1,2}$ and  $M$ are material-dependent parameters. In the Hamiltonian above we have neglected, without loss of generality, the momentum independent term proportional to the identity $\sigma_0 \otimes \tau_0$ since it corresponds to a rigid shift of all energies. In the $M B_{1,2} > 0$ inverted band regime, this model predicts the appearance of an idealized single surface Dirac cones thus verifying the ${\mathbb Z}_2=1$ non-trivial value of the strong topological index~\cite{Fu2007}. 
Since the bulk band gap $M$ in the stoichiometric materials is much larger than the energy scale of room temperature, external perturbations such as strain or pressure are not expected  to change the topology of the system, contrary to the case of, for instance, HgTe where a change from tensile to compressive strain drives a topological phase transition from a strong 3DTI to a Weyl semimetal phase~\cite{rua16}. 

However, the situation can be drastically different in a thin film where the surface Dirac cones located at opposite surfaces start to hybridize with each other. This hybridization, in fact,  yields a surface energy gap $\delta m \propto M e^{-w/l_s}$ where $w$ is the thin film thickness, while $l_s$ is the decay length of the topologically protected surface states~\cite{Liu2010}. Whenever the surface energy gap is  larger than a (low-temperature) thermal energy, the thin film then realizes a two-dimensional time-reversal invariant insulating state, which can be characterized by the so-called Fu-Kane-Mele ${\mathbb Z}_2$ topological invariant~\cite{Kane2005,Fu2006}. 
Furthermore, since the band gap is narrow, strain effects can and do lead to a change in the topology of the system. 
To prove the assertion above, we explicitly include strain effects in the effective bulk continuum Hamiltonian Eq.~\ref{H0}. Making use of the usual Bir-Pikus scheme~\cite{Brems2018}, and neglecting the possible presence of shear strains, the Hamiltonian for a generic biaxial strain takes the following form 
\begin{multline}
{\mathcal H}_{\mathrm{strain}}= [C_1 \epsilon_{zz} + C_2 (\epsilon_{xx}+\epsilon_{yy}) ]\sigma_0\otimes\tau_0 +\\+[ M_1 \epsilon_{zz}+ M_2 (\epsilon_{xx}+\epsilon_{yy})] \sigma_0\otimes\tau_3,
\label{Hstrain1}
\end{multline}
where $C_{1,2}$ and $M_{1,2}$ are the deformation potentials of the material at hand.

Although generally breaking the three-fold rotation symmetry ${\mathcal C}_{3v}$, homogeneous strains have a negligible effect even in thin film structures. 
This is immediately apparent considering the fact that besides a trivial rigid shift of the energy, a generic homogeneous strain merely renormalizes the bulk energy gap $M$ and consequently the thin film surface band gap $\delta m$ without affecting the topology of the system. 
However, and this is key, this does not hold true for inhomogeneous strains. 

\begin{figure}
\includegraphics[width=.9\columnwidth]{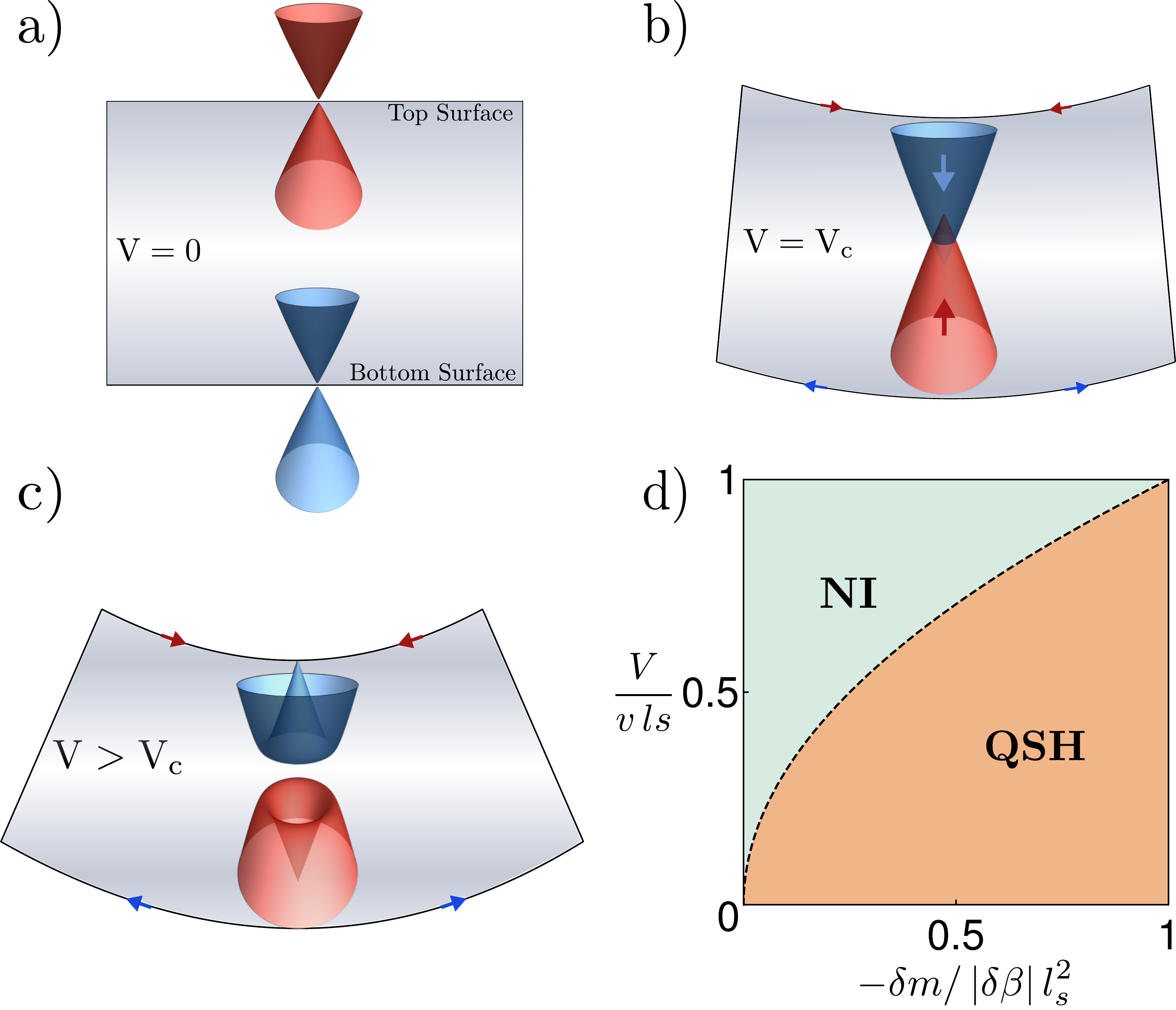}
\caption{(color online) Sketch of the surface band structure of a bent 3DTI thin film. a) In the strain-free configuration the top and bottom surface Dirac cones acquire an hybridization gap at the Dirac point. b) At a critical bending radius the two gapped Dirac cones are pulled toward each other and touch on a nodal line $|{\bf k}|=k_c$. c) For even larger bending radii the surface band gap is reopened with the system that then corresponds to a conventional band insulator. d) Corresponding topological phase diagram in terms of the strain-gradient induced structure inversion parameter $V$ and the ratio between the hybridization surface gap $\delta m$ and surface effective mass $\delta\beta$.}
\label{fig:fig1}
\end{figure}

Let us consider for instance the strain pattern due to a mechanical bending of the thin film. Such a mechanical deformation has been very recently achieved by placing single-crystalline Bi$_2$Se$_3$ nanowires over deep trenches~\cite{Schindler2017}. It can be also realized by external force loading using an atomic force microscope as recently employed to induce large flexoelectric effects in strontium titanate single crystals~\cite{Yang2018}.  As schematically shown in Fig.~\ref{fig:fig1}, in a mechanically bent thin film the top (bottom) surface is dilated while the bottom (top) surface is compressed. This, in turn, implies a structural inversion asymmetry that, as we show below, is decisive in tailoring the topological properties of the 3DTI thin film. 

\paragraph{Strain gradient-induced topological phase transitions --} 
To simplify our treatment, we will consider in the remainder an idealized bending with the local curvatures of the mechanical neutral plane [see Supplemental Material] that are constant. This produces 
a strain pattern $\epsilon_{xx}=\epsilon_{yy}=- \alpha z$ with $\alpha$ constant and  $z$ measured from the mechanical neutral plane, whereas $\epsilon_{zz}=-\nu \,  \epsilon_{xx} / (1-\nu)$ with $\nu$ the Poisson ratio. 
Hence, the strain Hamiltonian Eq.~\ref{Hstrain1} simply becomes 
\begin{equation}
\mathcal{H}_{\mathrm{strain}}(z)=  u_0\, \dfrac{z}{w} \sigma_0\otimes\tau_0 + u_3 \, \dfrac{z}{w} \sigma_0\otimes\tau_3, 
\label{Hstrain2}
\end{equation}
where we have introduced the two parameters $u_{0,3}$ that depend on the deformation potentials, the ratio between the bending radius and the thin-film thickness, and the Poisson ratio. 
To analyze the effect of such inhomogeneous strain, we first solve the unstrained Hamiltonian Eq.~\ref{H0} with open boundary conditions at the projected $\bar{\Gamma}$ point of the surface BZ. We therefore obtain two low-energy Kramers pairs $\ket{E^{\uparrow \, \downarrow}_{+}}$ and $\ket{E^{\uparrow \, \downarrow}_{-}}$ with opposite parity, whose energy  $m_\pm=\bra{E^{\uparrow \, \downarrow}_{\pm}} {\mathcal{H}}_{\mathrm{3DTI}}(0,0,-i\partial_z) \ket{E^{\uparrow \, \downarrow}_{\pm}}$ originates from the hybridization of the surface Dirac cones located at opposite surfaces. 
As the thin-film thickness is varied [see the Supplemental Material] these states undergo multiple crossings, which is consistent with the oscillatory crossover from 2DTI to 3DTI predicted in Ref.~\cite{Liu2010a}.
The inhomogeneous strain term couples equal spin states of opposite parity since, as mentioned above, it yields a structural inversion asymmetry.
It thus generates a nonvanishing coupling  $V=\bra{E^{\uparrow \, \downarrow}_{+}} {\mathcal{H}}_{\mathrm{strain}}(z) \ket{E^{\uparrow \, \downarrow}_{-}}$. 
This coupling, in turn, enhances the splitting of the low-energy states at the $\bar{\Gamma}$ point to $2 \sqrt{m^2 + V^2}$. However, it does not preclude the possibility of band-gap closing and reopening at different points of the surface BZ.

\begin{figure}
\includegraphics[width=1\columnwidth]{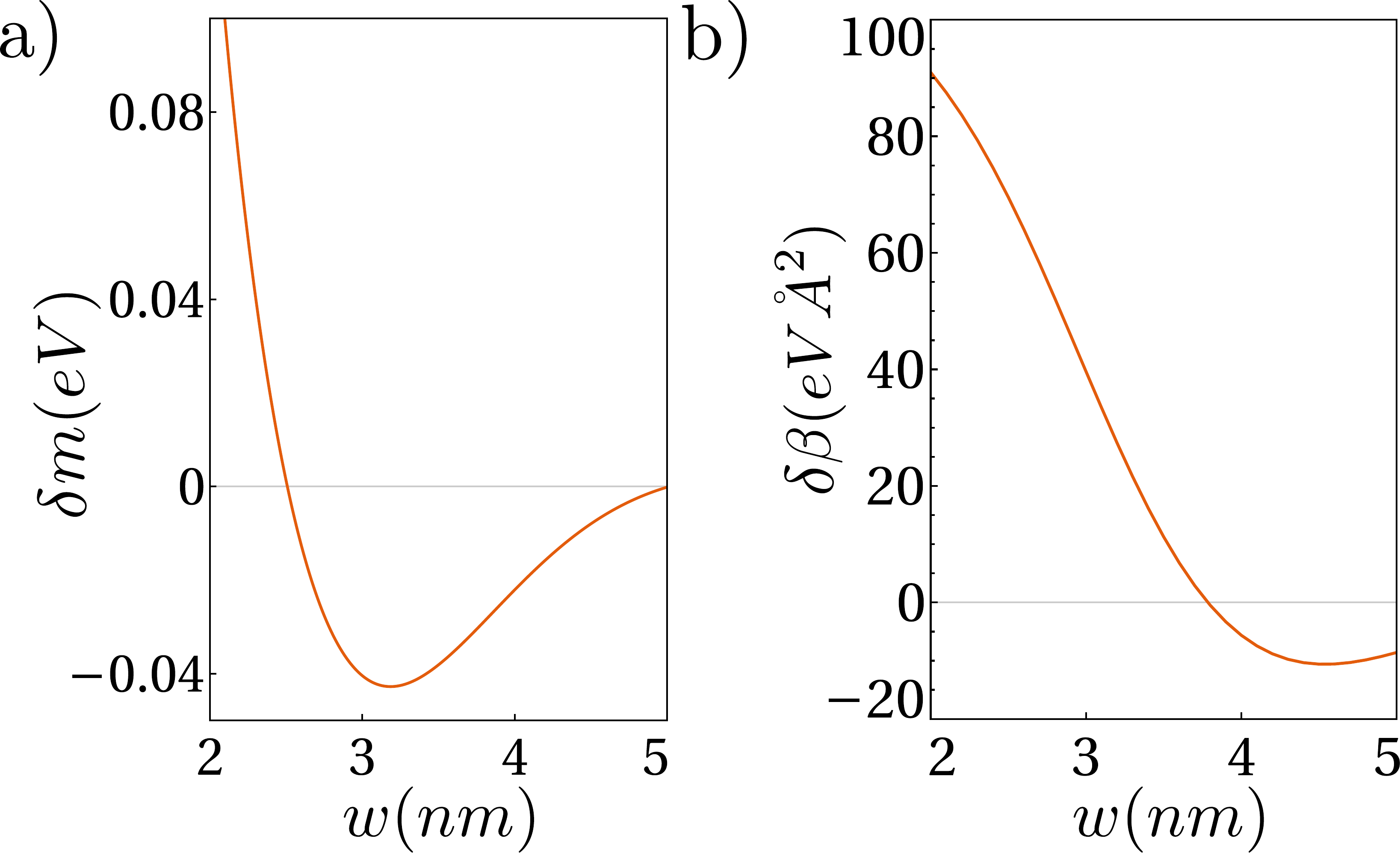}
\caption{(color online) Behavior of the surface Dirac cones hybridization gap $\delta m$ (a) and the effective mass parameter $\delta\beta$ (b) as a function of the 3DTI thin film thickness $w$. 
We have used the ${\bf k \cdot p}$ parameters of Bi$_2$Se$_3$ of Ref.~\cite{Zhang2009}.
For $2.5$~nm$\lesssim w \lesssim 5$~nm the unstrained thin film realizes a QSH insulator.}
\label{fig:fig2}
\end{figure}

To account for this, we next obtain the effective ${\bf k \cdot p}$ surface Hamiltonian in the $\ket{E^{\uparrow \, \downarrow}_{+,-}}$ subspace. It can be obtained by noticing that the parity eigenstates at the ${\bar \Gamma}$ point can be written as the bonding and anti-bonding states of the two surface Dirac states appearing in semi-infinite geometries, {\it i.e.} $\ket{E_{+,-}^{\uparrow \, \downarrow}} = ( \ket{t^{\uparrow \, \downarrow}} \pm \ket{b ^{\uparrow \, \downarrow}}) / \sqrt{2}$. In the $\ket{t^{\uparrow \, \downarrow}},\ket{b^{\uparrow \, \downarrow}}$ basis the surface Hamiltonian reads 
\begin{equation}
{\mathcal H}_{\mathrm{2D}}({\bf k})=
\epsilon(\mathbf{k})+
\begin{pmatrix}
V &  -i v k_+ & m(\mathbf{k}) & 0 \\
i v k_-& V & 0 & m(\mathbf{k}) \\
m(\mathbf{k}) & 0 & -V &  i v k_- \\
0 & m(\mathbf{k}) &  -i v k_+ & -V
\end{pmatrix}
\label{Hss}
\end{equation}
where $k_\pm=k_x\pm i k_y$ and we introduced the momentum dependent hybridization $m(\mathbf{k})=\delta m/2+\left |\delta\beta \right | (k_x^2+k_y^2)/2$, with $\delta m=(m_+-m_-) $, $\delta\beta=(\beta_+-\beta_-)$ and $\beta_{\pm}=B_2\bra{E_\pm^{\uparrow \, \downarrow}}\sigma_0 \otimes \tau_3\ket{E_\pm^{\uparrow \downarrow}}$. The particle-hole breaking term is defined as $\epsilon(\mathbf{k})=(m_++m_-)/2+(\beta_++\beta_-) (k_x^2+k_y^2)/2$ while the Fermi velocity $v=A_2\left |  \bra{E_\pm^{\uparrow \, \downarrow}}\sigma_1 \otimes \tau_1\ket{E_\mp^{ \downarrow\uparrow}}\right |$. For very thick films $\delta m \simeq 0$, we then recover two surface Dirac cones which, however, are pushed (pulled) to higher (lower) energies by the bending strain. This finding is perfectly compatible with low-temperature magnetotransport measurements of bent Bi$_2$Se$_3$ nanowires, which indeed show an opposite shift of the surface Dirac cones at opposite surfaces~\cite{Schindler2017}. More importantly, when expressed in the original $E^{\uparrow \, \downarrow}_{+,-}$ basis the surface ${\bf k \cdot p}$ Hamiltonian corresponds to the well-known Bernevig-Hughes-Zhang (BHZ) continuum model for HgTe quantum wells~\cite{ber06} but with the addition of a strain-gradient induced structure-inversion-asymmetry term $\propto V$. The evolution of the band structure while continuously increasing the structure inversion asymmetry term $V$ is shown in Fig.~\ref{fig:fig1} in the inverted band regime. 
For $V<V_c=v \sqrt{-\delta m/ \left |\delta \beta\right |}$ the system is adiabatically connected to a QSH insulator with a full surface band gap. By increasing the bending strain, a topological phase transition occurs at $V=V_c$ with the system that for $V>V_c$ becomes a conventional band insulator. Note that in the non-inverted regime the bending strain term $V$ does not lead to any band-gap closing and reopening points, which implies that inhomogeneous strains only act as trivializers of the system. 
In order to verify this physical picture, we have numerically computed [see Fig.~\ref{fig:fig2}] the hybridization gap $\delta m$ and the effective mass parameters $\delta\beta$ for thin films of varying thickness $w$ using the ${\bf k \cdot p}$ parameters for Bi$_2$Se$_3$ as obtained from density functional theory calculations~\cite{Zhang2009}. Using that $v \simeq A_2 \simeq 4$ eV $\AA$, we find that thin films of $2.6$~nm thickness can be driven via the topological phase transition for a strain-gradient induced coupling of $V \simeq 55$ meV. 
By considering that strain-induced shift of Dirac cones of $V \simeq 30$~meV have been achieved with 
a maximal strain at the top (bottom) surface of $\simeq \pm 0.1 \%$ ~\cite{Schindler2017},
one can therefore expect that the strain-gradient induced topological phase transition would require a maximal strain $\ll 1 \%$, or equivalently a bending radius of $\gg 260~$nm. 

The tunability of the topology in 3DTI thin films by means of inhomogeneous strains is similar in nature to the one achievable by means of externally applied electric fields, since they also provide a source of structure inversion asymmetry. It is thus natural to expect that when considering Cr-doped and V-doped Bi$_2$Te$_3$ thin films -- in the inverted band regime they gain considerable spin susceptibility through the van Vleck paramagnetism thereby developing long-range magnetic order with a quantized anomalous Hall effect~\cite{Yu2010} --  mechanical deformations will yield an analog of the Stark effect induced magnetic quantum phase transition theoretically predicted~\cite{Zhang2015} and experimentally verified~\cite{zhang2017} in these materials. 

\begin{figure}
\includegraphics[width=1\columnwidth]{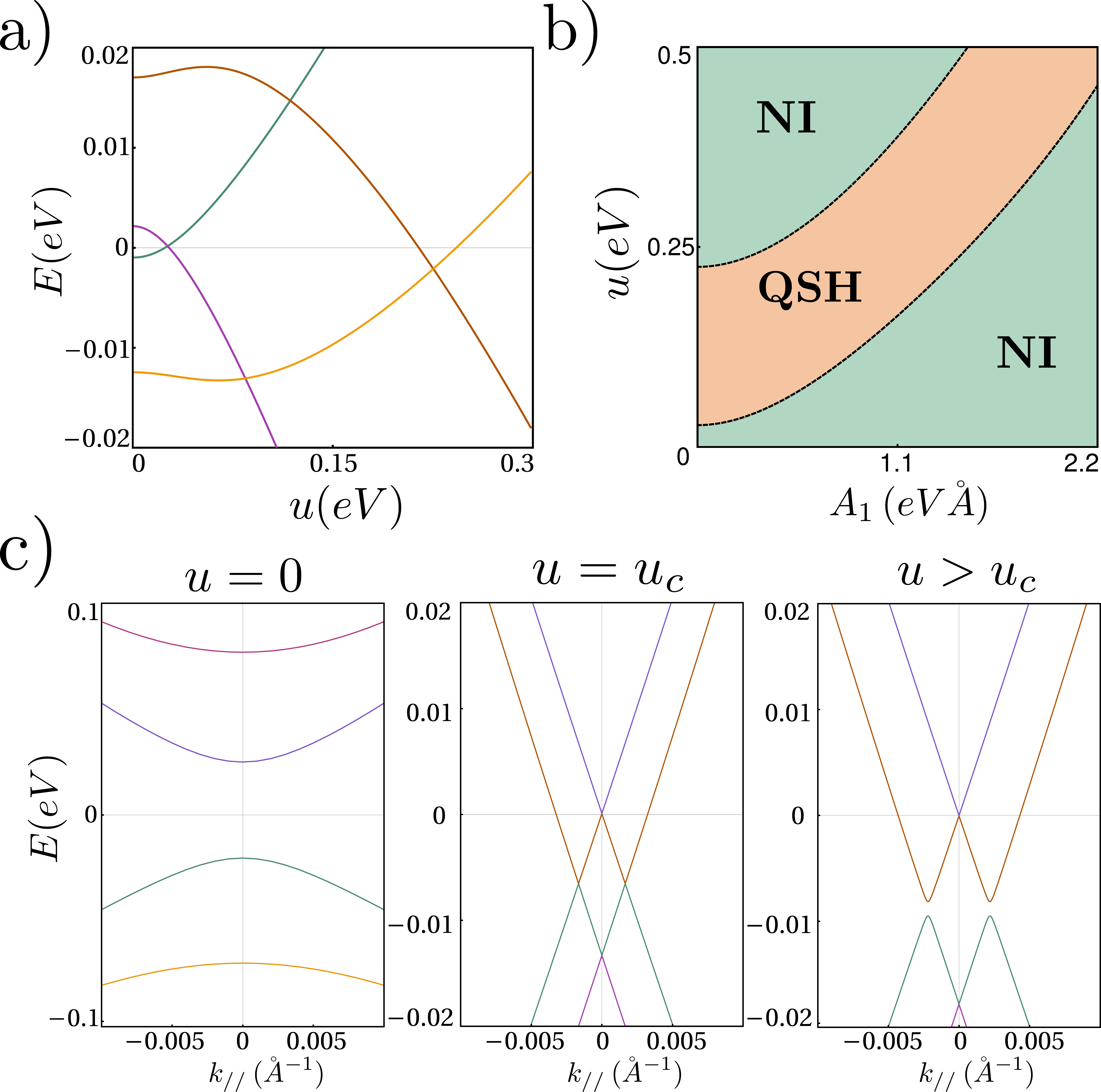}
\caption{(color online) a) Evolution of the $\bar{\Gamma}$ point energy levels for a narrow band gap 3DTI thin film with $A_1=0$ as the strain coupling terms are increased. The ratio between the strain coupling terms has been fixed to $u_0/u_3=0.2$. (b) Topological phase diagram in the $A_1$ vs strain plane. (c) Evolution of the band structure across the topological phase transition for $A_1 =2.2\, \mathrm{ev}\AA$.}
\label{fig:fig3}
\end{figure}

\paragraph{Narrow band-gap materials --} 
Having established the occurrence of strain-gradient induced topological phase transitions in materials where there is a large separation between the surface energy gap $\delta m$ and the bulk gap $M$, we next turn our attention to strong 3DTI that in their bulk are at the verge of a topological phase transition to a normal band insulator. In the Bi$_2$Se$_3$ material class, a substantial decrease in the bulk band gap can be obtained~\cite{Zhang2013} by increasing the Se content in Cr-doped Bi$_2$(Se$_x$Te$_{1-x}$)$_3$.  Similarly, density functional theory calculations~\cite{Zhang2010iop} predict a topological phase transition in Sb$_2$(Te$_{1-x}$Se$_x$)$_3$. 
Thin film structures of these alloys realize conventional band insulators due to a strong hybridization of the surface Dirac cones, in much the same way as ultra-thin films of Bi$_2$Se$_3$ but with thicknesses that can reach the tens of nanometers scale. In this case, however, the transversal subbands are very close to each other, which hence requires to go beyond the simple low-energy picture used so far. We have therefore numerically diagonalized the full Hamiltonian 
${\mathcal H}_{\mathrm{3DTI}}(\mathbf{k}) +{\mathcal H}_{\mathrm{strain}}(z)$ for a $15$~nm thick film 
considering a substantial decrease of the bulk gap to $M=2.8$~meV while starting from the special case $A_1 \equiv 0$. Fig.~\ref{fig:fig3}(a) shows the evolution of the $\bar{\Gamma}$ point  spin-degenerate subbands while increasing the strain couplings terms. The electron $\ket{E_n}$ and hole $\ket{H_n}$ subbands are not coupled by the strain gradients terms. Moreover, in each of the sectors the strain gradients realize an analog of the quantum-confined Stark effect~\cite{cendula2012} and thus push the electron (hole) levels downwards (upwards). Hence, at a critical strain coupling $u_c$ there is a crossing between the $\ket{E_1}$ and $\ket{H_1}$ subbands, which corresponds to a topological phase transition to a QSH phase. In fact, by projecting the full Hamiltonian onto these low-energy states, one finds that the effective Hamiltonian corresponds to the BHZ model with preserved two-dimensional ${\bf k}_{\parallel} \rightarrow -{\bf k}_{\parallel}$ inversion symmetry. 

Next we consider the effect of turning on the $A_1$ term. Since the crossing between $\ket{E_n}$ and $\ket{H_n}$ at the $\bar{\Gamma}$ are not protected by parity, they are changed to anticrossings for finite $A_1$ values [see the Supplemental Material]. However, by invoking the principle of adiabatic continuity the full band structure must still exhibit bandgap closing and reopening points which coincide with topological phase transitions. The ensuing topological phase diagram is shown in Fig.~\ref{fig:fig3}(b). 
Note that the band-gap closing reopening points occurs in this case on a nodal line ${\bf k}_{\parallel}=k_c$  as can be seen from the evolution of the band structure in Fig.~\ref{fig:fig3}(c). We finally note that this strain-gradient induced phase transition from normal insulator to QSH insulators can be also assisted via the application of an external electric field analogously to the electrical-induced topological transitions predicted to occur in few-layer phosphorene~\cite{liu2015}.

\paragraph{Conclusions --} 
To sum up, we have shown that bending strains couple to the massive Dirac quasiparticles of 3DTI thin films in such a way to trigger phase transitions between topologically distinct insulating phases. 
In stoichiometric materials with large bulk band gaps, the strain-induced structure inversion asymmetry generically endangers the parent QSH phase thus driving a topological phase transition to a normal band insulator.
In alloys with small bulk band gaps instead bending strains have a completely opposite effect, and can promote a topologically non-trivial phase as a result of the existence of a strain gradient analog of the quantum-confined Stark effect. The strain-gradient induced topological phase transition discussed in this work can also trigger magnetic quantum phase transitions in magnetic topological insulator thin films~\cite{Zhang2015,zhang2017} . This ``flexomagnetic" coupling could be employed in proposals for strain-assisted transistor devices as well as magnetic random access memories. 

\begin{acknowledgments}
C.O. acknowledges support from a VIDI grant (Project 680-47-543) financed by the Netherlands Organization for Scientific Research (NWO).
\end{acknowledgments}

\end{document}